\documentstyle[aaspp4,epsf]{article}
\begin{document}

\title{Properties of a Proper-Motion Selected Sample
of Giants in the Small Magellanic Cloud near NGC 121}
\author{Nicholas B. Suntzeff, Alistair R. Walker}
\affil{Cerro Tololo Inter-American Observatory}
\author{Verne V. Smith}
\affil{University of Texas at El Paso}
\author{Robert P. Kraft, Arnold Klemola}
\affil{UCO/Lick, UC Santa Cruz}
\author{Peter B. Stetson}
\affil{Dominion Astrophysical Observatory}
\vspace*{5mm}

We observed a sample of zero proper-motion stars ($\mu < 0.50 $ arcsec
cent$^{-1}$) from a field previously studied by Suntzeff et al.~(1986,
AJ, 91, 275) .  This field is 2.5\ arcdeg\ NW (2.6 kpc) of the center of
the SMC.  We obtained spectra for $\sim 40$ stars in the region of the
Ca~II infrared triplet using the CTIO Argus fiber-fed spectrograph.
We also obtained Argus echelle spectra of a single order at 6300\AA\
with $R=18000$ in one run.  The low-dispersion spectra were reduced to
metallicities based on the Ca II equivalent widths using the Da Costa
\& Armandroff (1995, AJ, 109, 2533) technique and the metallicity
scale from Zinn \& West (1984, ApJS, 55, 45). The typical abundance
error is 0.12 dex.  For half the sample, we have echelle velocities
which are accurate to 1.5 km s$^{-1}$. For the rest of the sample, the
low-dispersion data yield single-observation velocities accurate to
about 5 km s$^{-1}$ based on repeat observations.

The average properties of the sample are:

\begin{center}
$\overline{{\rm [Fe/H]}} = -1.33,\ \sigma = 0.42,\ {\rm N}=35$\\
$\overline{\rm v} = 135.4\ {\rm km\ s^{-1}},\ \sigma = 25.7\ {\rm km\ s^{-1}},\ {\rm N}=36$\\
\end{center}

We compare these results to the data tabulated in Da Costa \&
Hatzidimitriou (1998, AJ, 115, 1934):

\begin{center}

\begin{tabular}{lccc} 
\multicolumn{4}{c}{\bf Kinematics of SMC populations}\\ 
\hline \hline
\multicolumn{1}{c}{Sample}
&\multicolumn{1}{c}{$\overline{\rm v}$}
&\multicolumn{1}{c}{$\sigma$}
&\multicolumn{1}{c}{N}\\
&\multicolumn{1}{c}{ km s$^{-1}$} 
&\multicolumn{1}{c}{ km s$^{-1}$} \\
\hline
This sample    & 135.4  & $25.7 \pm 2$  & 36 \\
PN            & 142    &  $25 \pm 3$   & 44 \\
C stars       & 148    & $27 \pm 2$   & 131 \\
Star clusters & 138    &  $16 \pm 4$  &  7 \\
Cepheids      & complicated & $22 \pm 3$   &  61 \\
HI            & \nodata & $25 \pm 0.6$ & \nodata \\
\hline \hline
\end{tabular}
\end{center}

\begin{figure}[ht]
\epsscale{0.5}
\plotone{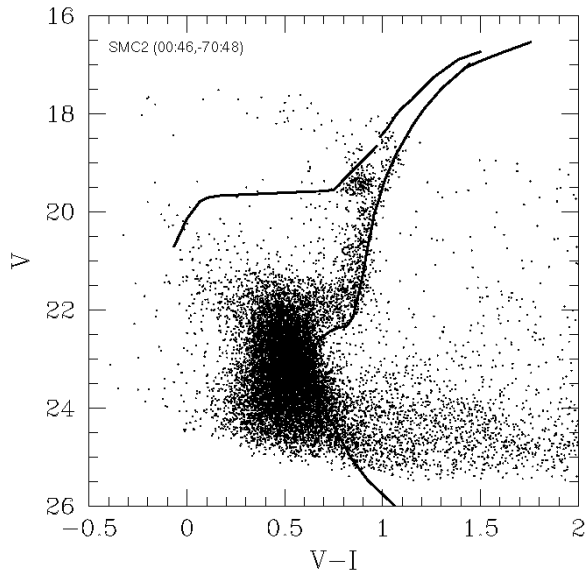} 
\vspace*{-8mm}
\caption{Color magnitude diagram of the SMC field and the ridge line
for M5 moved to the distance and reddening of the SMC.} \label{f1}
\end{figure}

In Figure \ref{f1} we plot a $VI$ cmd from CTIO 4m data. The ($V-I$)
zero point is provisional. We also plot the M5 ridge line from
Sandquist et al.~(1996, ApJ, 470, 910) and Johnson \& Bolte (1998, AJ,
115, 868).  The M5 ridge line has been moved to the distance of the
SMC using the relative $V$ magnitudes of the RR Lyraes in M5 and the
field of the SMC adjusted for reddening differences. No metallicity
correction is needed.  The M5 ridge line forms the lower envelope of
the subgiant branch (SGB). The width of the SGB in $V$ seems to
indicate that star formation proceeded for about \mbox{$\sim6$ Gyrs}
after the corresponding age of M5, and then essentially stopped. If
the most metal-rich stars are the youngest (which we have not shown),
then the field metallicity stopped at about \mbox{[Fe/H]=--0.5}. From the lack
of a well defined HB and given the low mean metallicity, there is no
evidence for a populous old component.

\vspace*{3mm}
The summary of our results:

\begin{itemize}

\item The mean metallicity of the SMC field near NGC 121 is ${\rm
[Fe/H]}=-1.3$ with a real dispersion of 0.4 dex. The most metal rich
stars are at --0.5 dex and the most metal-poor stars at --2.1 dex in a
sample of 35 stars.

\item The velocity dispersion of the sample is 25 km s$^{-1}$.  The
dispersion of various kinematic samples in the SMC is independent of
age.

\item The RGB colors are consistent with Galactic globular clusters of
similar metallicity. The field population, however, does not have an
extended horizontal branch, and the main sequence turnoff extends 1
magnitude brighter than the turnoff in M5. Evidently {\it active} star
formation in this region started roughly at the age of M5 and extended
for about 6 Gyrs, and then stopped.

\end{itemize}




\end{document}